\documentclass[aps,twocolumn,showpacs,superscriptaddress]{revtex4}
\usepackage{amsmath,amssymb}
\usepackage{graphics,graphicx}
\usepackage{dcolumn,bm}
\newcommand{\cpt}
{\affiliation{Centre de Physique Th\'{e}orique (CNRS UMR 6207), Universit\'{e} de la M\'{e}diterran\'{e}e Aix Marseille II,
Luminy, 13288 Marseille cedex 9, France.}}
\newcommand{\cu}
{\affiliation{Department of Physics, University of Calcutta, 
92 Acharya Prafulla Chandra Road, Kolkata 700009, India.}}

\begin{document}

\title
{Antipersistent dynamics  in  kinetic models of wealth exchange}

\author{Sanchari Goswami}%
\email[Email: ]{sg.phys.caluniv@gmail.com}
\cu
\author{Arnab Chatterjee}%
\email[Email: ]{arnab.chatterjee@cpt.univ-mrs.fr}
\cpt
\author{Parongama Sen}%
\email[Email: ]{psphy@caluniv.ac.in}
\cu

%\email{psphy@caluniv.ac.in}

\begin{abstract}
We investigate the detailed dynamics of gains and losses  made by agents in  some kinetic models of
wealth exchange. 
% where
%agents have a saving propensity $\lambda$ distributed randomly ($0 \leq \lambda < 1$). 
An earlier work suggested that a walk in an abstract gain-loss space can be conceived for 
the agents. For models in which agents do not save, or save with uniform saving propensity,
the walk has diffusive behavior. In case the   saving propensity $\lambda$ is distributed  randomly ($0 \leq \lambda < 1$),
the resultant walk  showed a  ballistic nature (except at a particular value of  
$\lambda^* \approx 0.47$).  Here we consider several other 
features of the walk with random $\lambda$. 
While some macroscopic properties of this walk are comparable to a 
biased random walk, at microscopic level, there are gross differences.
The difference turns out to be due to an antipersistent tendency  
  towards making a gain (loss) immediately after making a loss (gain). 
This correlation is in fact present in kinetic models without saving or 
with uniform saving as well, such that the corresponding walks are not 
identical to ordinary random walks. In the distributed saving case, 
antipersistence occurs with a simultaneous overall bias.

\end{abstract}

\maketitle
\section{Introduction}
The distribution of wealth $P(m)$ among individuals in an
economy largely show an universal pattern, decaying as $P(m) \sim m^{-(1+\nu)}$
for large values of wealth $m$. $\nu$ is called the Pareto exponent~\cite{Pareto:1897},
and is usually between $1$ and $3$~\cite{Mandelbrot:1960,EWD05,ESTP,SCCC,Yakovenko:RMP,datapap}. 
A number of models have been proposed in recent times to 
 reproduce these observed features, specifically to obtain a power law tail as was observed in empirical data.
Some of these models have been inspired by the kinetic theory of gas-like exchanges, where
a pair of traders exchange wealth, respecting local conservation in any trading
\cite{marjitIspolatov,Dragulescu:2000,Chakraborti:2000,Chatterjee:rev,Chakrabarti:2010,Chatterjee:2010}.
These models have a microcanonical description and nobody ends up with negative wealth (i.e., debt is not allowed).
Thus, for two agents $i$ and $j$ with money $m_i(t)$ and $m_j(t)$ at time $t$, the general trading process is given by:
\begin{equation}
\label{mdelm}
m_i(t+1) = m_i(t) + \Delta m; \  m_j(t+1) = m_j(t) - \Delta m;
\end{equation}
time $t$ changes by one unit after each trading.

In a simple conservative model proposed by  Dragulescu and Yakovenko (DY model) \cite{Dragulescu:2000},  $N$ agents 
exchange wealth or money randomly keeping the total wealth $M$ constant. 
The steady-state ($t \rightarrow \infty$) wealth follows a Gibbs distribution:
$P(m)=(1/T)\exp(-m/T)$; $T=M/N$, 
a result which is robust and independent of the topology of the (undirected)
exchange space~\cite{Chatterjee:rev}.

An additional concept of \textit{saving propensity} was considered first by Chakraborti and Chakrabarti~\cite{Chakraborti:2000} 
(CC model hereafter). Here, the agents save a fixed fraction $\lambda$ of their wealth
when  interacting  with another agent. Thus,
%Two agents with initial money $m_i(t)$ and $m_j(t)$ at time $t$ interact such that 
%they end up with money $m_i(t+1)$ and $m_j(t+1)$
%given by
\begin{equation}
\label{fmi}
m_i(t+1)=\lambda m_i(t) + \epsilon_{ij} \left[(1-\lambda)(m_i(t) + m_j(t))\right],
\end{equation}
\begin{equation}
\label{fmj}
m_j(t+1)=\lambda m_j(t) + (1-\epsilon_{ij}) \left[(1-\lambda)(m_i(t) + m_j(t))\right];
\end{equation}
$\epsilon_{ij}$ being a random fraction between $0$ and $1$, modeling the stochastic nature
of the trading.
It is easy to see that the $\lambda=0$ case is equivalent to the DY model.
This results in completely different types of wealth 
distribution curves, very close to Gamma distributions~\cite{Patriarca:2004,Repetowicz:2005,Lallouache:2010}
which fit well to empirical data for low and middle wealth regime~\cite{datapap}.
The model features are somewhat similar to Angle's work~\cite{Angle}.
Obviously, the CC  model did not lead to the expected behaviour according to Pareto law.

In a later  model  proposed by Chatterjee et. al.~\cite{Chatterjee:2004} 
(CCM model hereafter) it was assumed that the saving propensity has a distribution
and this immediately led to a wealth distribution curve 
with a Pareto-like tail. Here, 
\begin{equation}
\label{mi}
m_i(t+1)=\lambda_i m_i(t) + \epsilon_{ij} \left[(1-\lambda_i)m_i(t) + (1-\lambda_j)m_j(t)\right],
\end{equation}
\begin{equation}
\label{mj}
m_j(t+1)=\lambda_j m_j(t) + (1-\epsilon_{ij}) \left[(1-\lambda_i)m_i(t) + (1-\lambda_j)m_j(t)\right];
\end{equation}
which are different from the CC model equations as $\lambda$'s are now agent dependent.
Various studies on the CCM model have been made 
soon after~\cite{Chatterjee:2005,Mohanty:2006,Kargupta,ecoanneal,Toscani,Chatterjee:2009,ChakrabartiASBK,Chakraborty:2010}. 

In a recent study,  the agent dynamics for models
with saving propensity was studied  with emphasis on the nature of transactions (i.e., whether it
is a gain or a loss)~\cite{Chatt-sen:2010}. It was observed  that in the CCM model, 
 the amount of money gained or lost  by a tagged agent in a single interaction follows a distribution  which is not symmetric in general,
well after equilibrium has reached. 
The distribution strongly depends on the saving propensity of the agent. For example, an agent with larger $\lambda$ suffers more losses 
of less denomination compared to an agent with smaller $\lambda$ 
 although the total money of the two agents has reached equilibrium, that is, each agent's money fluctuates around  a 
$\lambda$ dependent value. 

In \cite{Chatt-sen:2010}, in order to study the dynamics  of the transactions (i.e., gain or loss),  a walk 
was  conceived  for the agents in an abstract one dimensional gain-loss space (GLS) where the agents  conventionally
take a step towards right if a gain is made and left otherwise.  
It was  found that for this walk, $\langle x\rangle$, the distance travelled scales linearly with time $t$ suggesting a ballistic 
nature of the walk for the CCM walk. Moreover,  the slope of the  $\langle x\rangle$ versus $t$ curves is dependent on $\lambda$; it    is positive 
for small $\lambda$ and continuously goes to negative  values for larger values of $\lambda$. The slope becomes zero at
a value of $\lambda^* \simeq 0.469$.  In general for the CCM walk $\langle x^2 \rangle$ scales with $t^2$ .
For the CC model on the other hand, $\langle x^2 \rangle$ scaled with $t$ as in a random walk while $\langle x \rangle  \approx 0$.
%In fact, the scaling of $\langle x^2\rangle$ shows an initial variation 
%as $t$ before crossing over to the $t^2$ variation.
% and the time scale over which the linear variation is valid 
%increases as one approaches $\lambda^*$.

The above results na\"{i}vely suggests that the walk in the GLS is like a biased random walk (BRW) (except perhaps at $\lambda^*$) for the CCM model while it is like a random walk (RW) for the CC model.
In fact, in the CCM model, associated with each value of $\lambda$,  there seems to be a unique value of the parameter $p$ 
characterizing the corresponding biased random walk, where $p$ is 
is the probability of moving towards a particular direction.   
This makes it convenient to compare the CCM walk with a BRW which we discuss  
%We first discuss the comparison of the CCM walkers with 
%conventional biased random walkers 
in the next section by considering some additional features of the walk. 
The  results lead to a study of the temporal correlations presented in Sec.~\ref{sec:3}.
In Sec.~\ref{sec:4} we discuss the results of another walk, the simulated walk (SW),
 which can be generated in the GLS using the results 
of \cite{Chatt-sen:2010} to make the  analysis more conclusive. 
In the last section the results are summarized and discussed.

\section{CCM walk in the GLS: comparison with BRW}
\label{sec:2}

Our aim is to compare the results of the walk in GLS in the CCM model with those of a  BRW in this section. 

In a BRW, a walker moves towards a particular direction with probability $p \neq 1/2$ such that 
the total distance $\langle x \rangle$ travelled is linear in time $t$, precisely 
$\langle x \rangle = (2p-1) t$ in the preferred direction. 
To compare the CCM walk with the BRW we have the following scheme: 

First, we extract  effective values of $p$  for the walk in the GLS using the slopes of the 
$\langle x \rangle$ versus $t$ plots assuming it is a 
BRW. Next, from the distribution of distances travelled without change in direction in the CCM walk, we again
extract effective $p$ values assuming it is a BRW.   

We also compare the direction reversal probability of the CCM walk to that of the BRW. If these 
effective values of $p$ and direction reversal probability of the CCM walk and the BRW turn out to be identical, 
 we can  conclude that the walks in the GLS for the CCM model are ordinary biased
random walks. 

\subsection{$p$  using slopes of  $\langle x \rangle$ versus $t$ curves}

\begin{figure}%[t]
\includegraphics[width=8.5cm]{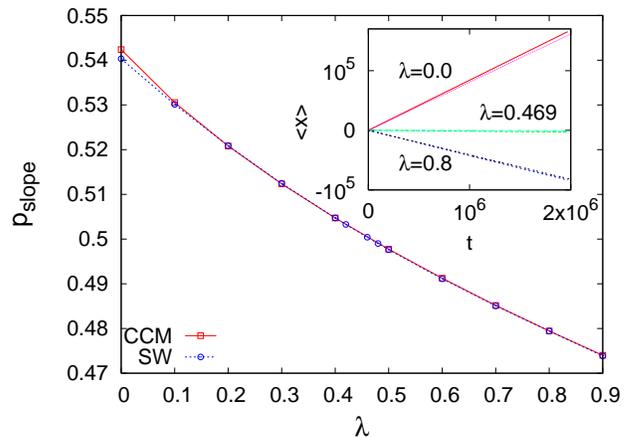}
\caption{(Color online) Plot of  $p_{slope}$ as a function of $\lambda$ obtained from the 
slopes of 
$\langle x \rangle$ versus $t$ plot for the CCM model (section II) for $N= 256$  and the 
simulated walker (SW, discussed in section IV). Inset shows  that the variation of  
 $\langle x \rangle$ against $t$ for 
$\lambda = 0.0, 0.469$ and $0.8$ for the 
 CCM model  and the SW are almost indistinguishable. 
}
\label{p-slope}
\end{figure}

As already mentioned, for the CCM model,  $\langle x \rangle$, the distance travelled  in time $t$ varies linearly with $t$. 
Thus, $\langle x \rangle = s_{0}t$ and an effective $p \equiv p_{slope}$ can be calculated using the 
relation  $p_{slope} = \frac{s_{0}+1}{2}$.
 (By our convention, if $p_{slope}> \frac{1}{2}$, the walker has a bias towards right (gain).) 
The results obtained in this way are shown in Fig.~\ref{p-slope}. 
We notice that  $p_{slope}$ approaches   $\frac{1}{2}$  as $\lambda \to \lambda^*$.
%%%%%%%%%%%%%%%%%%%%%%%%%%%%%%%%%%%%%%%%%%%%%%%%%%%%%%%%%%%%%%%%
%********************************************************

\subsection{Distribution of distances travelled without change in direction}

\begin{figure}%[t]
\includegraphics[width=6.0cm,angle=270]{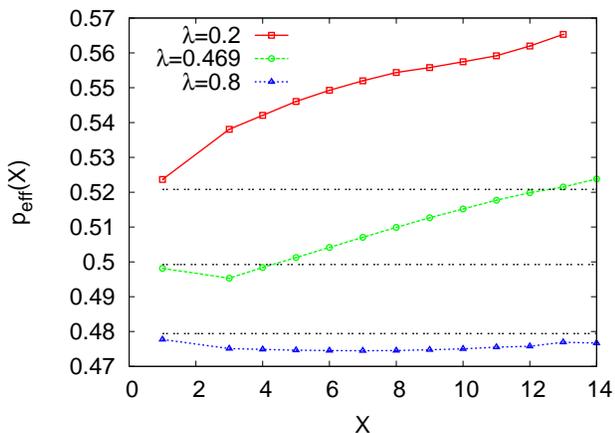}
\caption{(Color online) Variation of $p_{eff}(X)$ against $X$ for $\lambda =  0.2, 0.469, 0.8$ for the CCM model.
$p_{eff}$ values are  not  independent of $X$ in general. The horizontal lines from top to bottom 
indicate the values of $p_{slope}$ corresponding 
to $\lambda =  0.2, 0.469, 0.8$ for the CCM model.
}
\label{p-vs-X-CCM}
\end{figure}

We  study the distribution of the walk lengths $X$ through which the walker travels without any change in direction.
For the BRW, this is easy to calculate:  in our convention let the  probability to move towards right be $p$, 
then the probability $W_s(X)$ that a walker goes through a length $X$ at a stretch along the right  direction 
is proportional to $p^X (1-p)^2$. The corresponding probability along left is written as $W_s(-X) \propto    
 p^2 (1-p)^X$, and therefore,  in a BRW,
\begin{equation}
\frac{W_s(X)}{W_s(-X)} = \left(  \frac{p}{1-p} \right) ^{X-2}.
\label{ratio}
\end{equation}
For the walk in the GLS,  $W_s(X)/W_s(-X)$ is calculated numerically 
for any value of $\lambda$, and  a  value of $p_{eff}(X, \lambda)$ for
different values of $X$ is obtained using the above equation.
 If the CCM walkers were really  simple biased random walker, one would get a $p_{eff}(X, \lambda)$
{\it independent} of $X$ for a given $\lambda$ and close to the value  $p_{slope}$ obtained using the slope method.
 In Fig. \ref{p-vs-X-CCM},  we plot the $p_{eff}(X)$.
It should be noted that in this method, $p_{eff}(X=2)$ cannot be obtained as the R.H.S. of Eq.~\ref{ratio} becomes unity, i.e., 
$p$ independent. 
We notice immediately that the effective $p$ values are in no way independent of $X$ (except perhaps when $\lambda$ is
close to unity). 
This strongly indicates that
the walks are not simple BRW. 
We will get back to this issue in Sec.~\ref{sec:4} again.

%%%%%%%%%%%%%%%%%%%%%%%%%%%%%%%%%%%%%%%%%%%%%%%%%%%%%%%%%
%%%%%%%%%%%%%%%%%%%%%%%%%%%%%%%%%%%%%%%%%%%%%%%%%%%%%%%%%%

\subsection{Probability of direction reversal}

Another quantity closely related to the measure discussed in the previous subsection  is the probability of direction reversals made by the 
walker, which is defined as $f = n_d/n$, where $n_d$ is the number of times the walker changes direction 
and $n$ the total number of steps  (duration of the walk).  $f$ can be identified as $1/\langle X \rangle$, where  
\begin{equation}  
\langle X \rangle = \sum_X \big[ X W_s(X) +X W_s(-X)\big]
\label{avX+}
\end{equation}
is the average distance travelled at a stretch.
 Note that we have 0 the probabilities $W_s(X)$ such that ${\sum_X [W_s(X) + W_s(-X)]}=1$.
%********************************************************
\begin{figure}[ht]
\includegraphics[width=8.5cm]{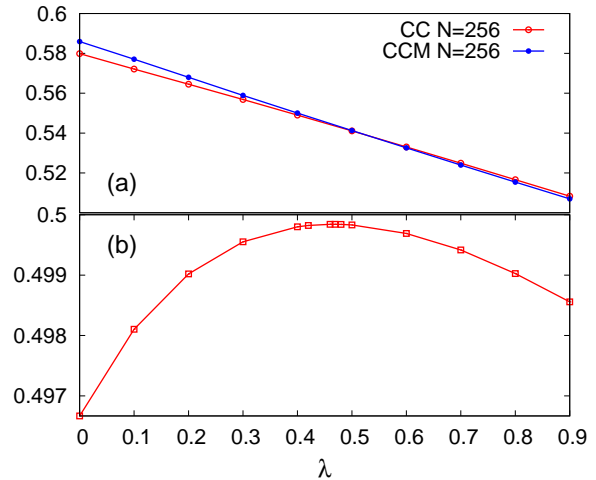}
\caption{(Color online) (a) Plot of direction reversal probability $f$ against $\lambda$ for CCM and CC models. (b) Same for the simulated walk.}
%with either the RBW data or the original BRW data.
%}
\label{direc-change}
\end{figure}
%%%%%%%%%%%%%%%%%%%%%%%%%%%%%%%%%%%%%%%%%%%%%%%%%%%%%%%%%%%%

The  probability  
of direction reversal for the BRW   is $2p(1-p)$  and has a maximum value  of  $f = 1/2$ at $p=1/2$ which corresponds to a random walk.
However, we get the result that  for the CCM model, $f$ is always greater than 1/2.  The data is shown in Fig \ref{direc-change}. 
Thus there is no way one can extract an equivalent value
of $p$ and make comparisons. 
This again shows that the agents in the CCM model   do not perform   a biased random walk in the 
gain loss space. 

\begin{figure}%[t]
\includegraphics[width=8.5cm]{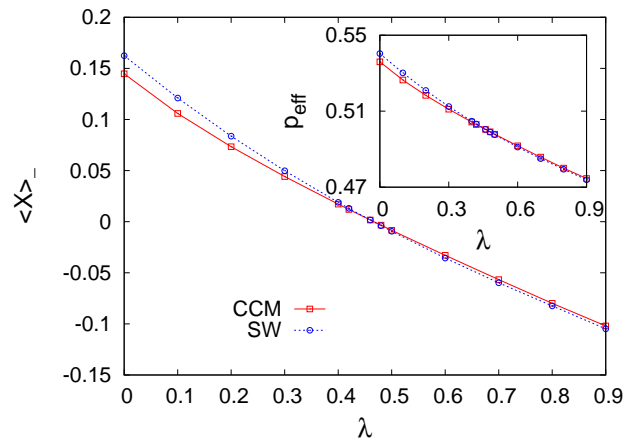}
\caption{(Color online) $\langle X \rangle_{-}$  plotted against $\lambda$ for both the 
CCM model (section II) and SW (section IV).
Inset shows the effective $p$ values using $\langle X \rangle_{-}$. 
}
\label{avXcompare}
\end{figure}

One can also define a quantity 
\begin{equation}
 \langle X\rangle_{-} = \sum_X \big[ X W_s(X) - X W_s(-X)\big]
\label{avX-}
\end{equation}
to   obtain
 an effective $p$ value for each $\lambda$  using  the fact that for the BRW, 
$\langle X \rangle_{-} = (2p-1)/(2p(1-p))$. 
$\langle X\rangle_{-}$ is
shown as a function of $\lambda$ in Fig.~\ref{avXcompare}.
Interestingly, here it is possible to extract effective values of $p$ which are quite close to $p_{slope}$, the values  
obtained using the slope method (data shown in Fig.~\ref{avXcompare} to be compared with the data in Fig.~\ref{p-slope}).  
%Clearly,  $\langle X\rangle_{-}$ is a coarse grained  
%quantity, which erases out the 
%details due to  which the 
%differences occur in $W_s(X)$.
%**********************************************************

%***********************************************************

Thus we find that the results for   $f$ (which is related to $\langle X\rangle$),  indicate that the CCM walk on the 
GLS cannot be regarded as  a BRW while the measure $\langle X\rangle_{-}$ is  fairly consistent with it.
.
In the following sections we resolve this intriguing issue.

Before ending this section, we make a few comments about the quantities $\langle X \rangle$ and $\langle X \rangle_{-}$, the first of which is directly 
related to the direction reversal probability $f$. 
If the  left and right moves of a walker are regarded as the states of a Ising spin and the temporal sequence of 
the moves are viewed as spin states of the consecutive sites of a  one  dimensional lattice, 
 then $\langle X \rangle$ is equivalent to the average domain size and  $\langle X\rangle_{-}$ can be interpreted as
the magnetization. 
Also,  it should be noted that $\langle X\rangle_{-}$ is a quantity which will be zero if the $W_s(X)$ distribution is symmetric.

\section{Correlations}
\label{sec:3}

Earlier, we had mentioned that for the CC model, the walkers apparently behave as ordinary random walkers.
Since the probability of direction reversal in the CCM model shows drastic difference  when compared to 
BRW, the question whether $f$ is exactly equal to $1/2$ in the CC model (as in a random walk) can also be raised. 

 Interestingly, the CC model shows little difference with the CCM 
when $f$  is compared (Fig. \ref{direc-change})  while previous  studies had shown that the scaling of $\langle x^2 \rangle$ are 
 quite different for the two models.
We are therefore led to investigate more into the 
details of the walks for both the CC and CCM walks in the context of direction changes.

 In the CCM model, $f > 1/2$  while a bias depending on 
 In the CCM model, $f > 1/2$  while a bias depending on 
$\lambda$ is simultaneously  maintained.  It may seem a little difficult to conceive such a walk, but it is possible to 
construct some deterministic toy walk models which have these properties. 
For example, a walk  which goes along right (R) and left (L) as  RRLRRL etc.~\cite{manna},  
has these features. Here there is a overall bias 
towards right while $f = 2/3 > 1/2$. Adding some noise may still maintain 
the bias and  $f > 1/2$.
%********************************************************
\begin{figure}%[t]
\includegraphics[width=8.5cm]{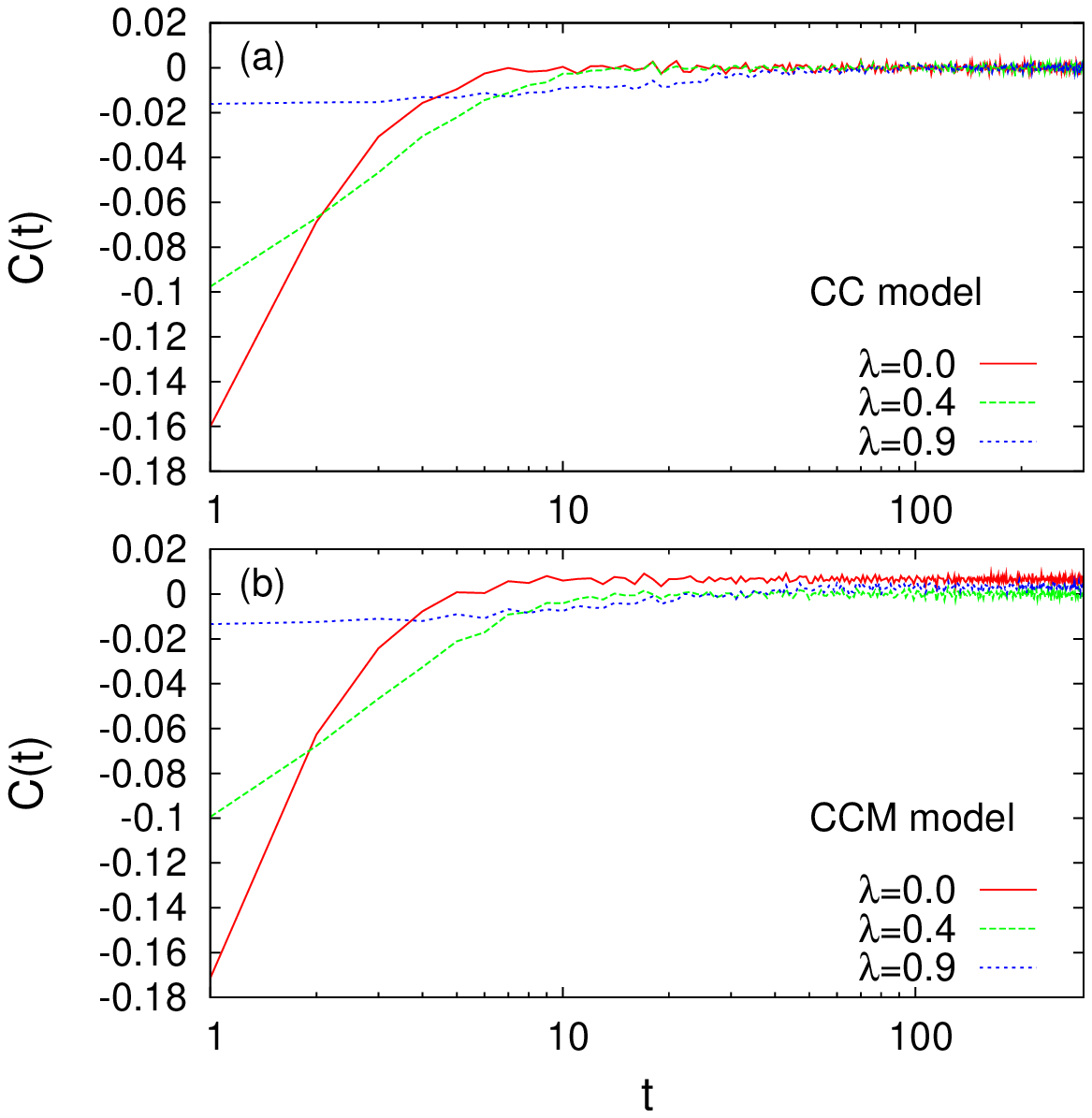}
\includegraphics[width=8.5cm]{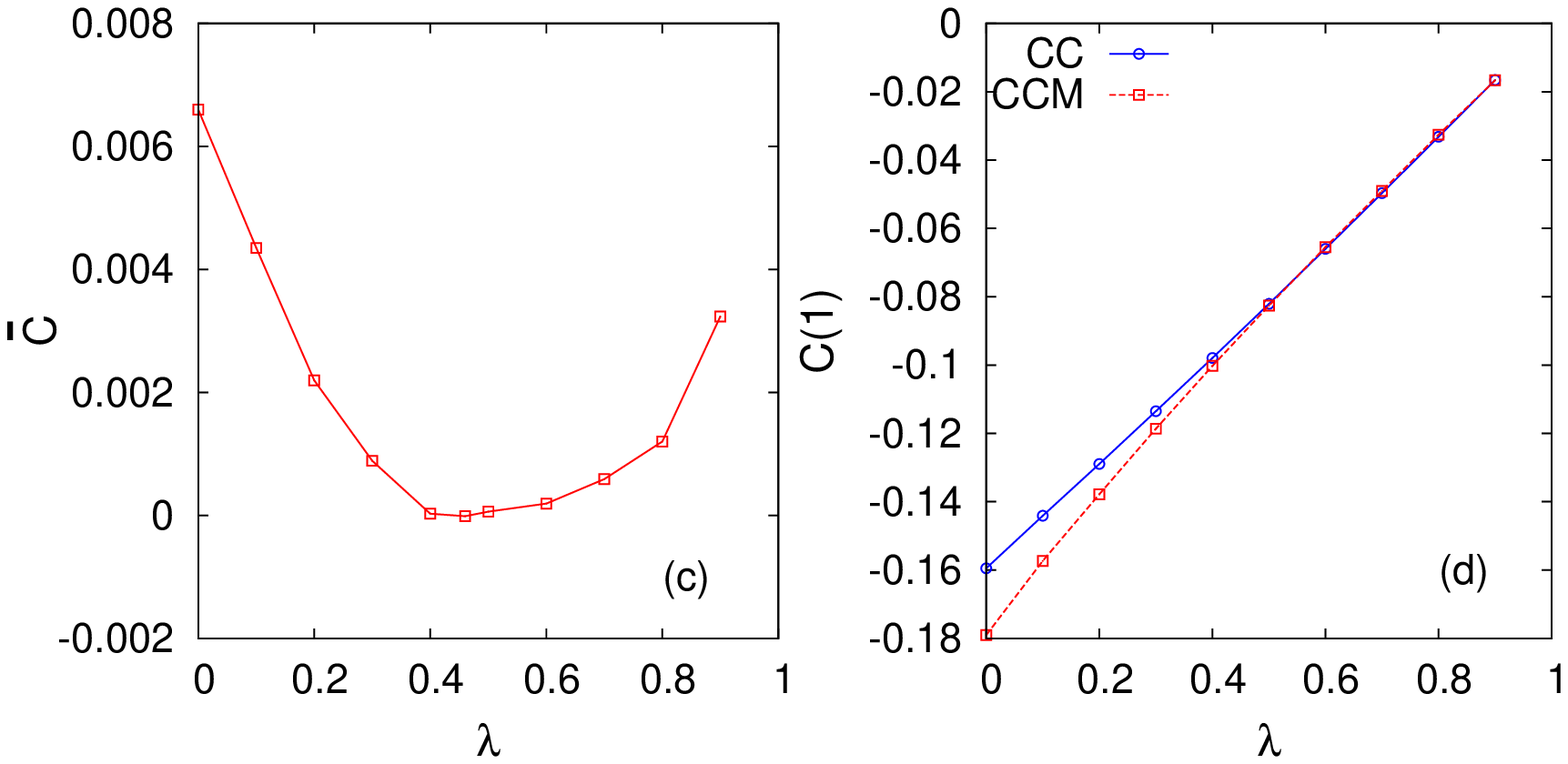}
\caption{(Color online) 
(a) The correlation of steps taken at  time intervals of $t$  are calculated
using  Eq.~\ref{corr-define} for a single walk configuration for CC  model,
averaging over all possible initial times $t_0$. (b) Same for the CCM model.
(c) Saturation value of the correlation, $\bar C$, 
at long times for the CCM model shows a dependence on  $\lambda$; it is $\approx 0$ at 
$\lambda^*$ and increases as $\lambda$ deviates from $\lambda^*$. In (d),
the correlation for two consecutive time steps, $C(1)$, is shown for both
the CC and CCM models also as a function of $\lambda$.} 
\label{corr-time}
\end{figure}
%%%%%%%%%%%%%%%%%%%%%%%%%%%%%%%%%%%%%%%%%%%%%%%%%%%%%%%%%%
The CC walkers on the other hand also show a deviation from a simple 
random walk as $f > 1/2$ is  obtained here.

Since a large value of  probability of  direction changes implies  that there is a higher probability 
of taking two successive steps in directions opposite to each other, 
it immediately suggests that there is a correlation between successive steps. Let the  step taken  at time $t$ be written as  
 $s(t) 
=  \pm 1$ ($+1$ for a right step and $-1$ for a left step). 
The time correlation function $C(t)$ is then defined as 
\begin{equation}
C(t) = \langle s(t_0)s(t_0+t)\rangle  - {\langle s(t_0) \rangle}{\langle s(t_0+t) \rangle}, 
\label{corr-define}
\end{equation}
where $t_0$ is an arbitrary time after equilibrium has reached.  

We take average over different initial times $t_0$ to calculate the 
above correlation in a single realization of a walk for both the CC and CCM models. 
The second term on the R.H.S. of Eq. \ref{corr-define} can be replaced 
by $s{_0}^2$, as 
 ${\langle s(t_0) \rangle}$,   the average step length,  is independent of time at equilibrium and equivalent to 
 $s_0$, the slope  of the ${\langle x \rangle}$ versus $t$ plot. 
For the CC walk,  therefore, 
${\langle s(t_0) \rangle}  = 0$ while for the CCM walk it has a nonzero value.
We notice that for both CC and CCM walks, there is a strong correlation when $t=1$, 
which decays quite fast for both models.   
For the CC walk, the correlations become zero 
at later times 
  (Fig \ref{corr-time}). 
For the CCM model, however, the correlation saturates to a very small nonzero
 value
 which is $\lambda$ dependent. The saturation value $\bar C = C(t \rightarrow \infty)$ is estimated by 
averaging $C(t)$ over the last few hundred steps. The average saturation values $\bar C$ are shown 
in the inset of Fig \ref{corr-time} as a function of $\lambda$. $\bar C$ has a minimum value $\sim O(10^{-5})$
close to $\lambda^*$ and a small positive value which increases as $\lambda$ deviates from $\lambda^*$.

The short time correlation in both models is indeed 
negative which is consistent with the fact that direction reversal
occurs with a probability $> 1/2$. It may be mentioned that 
for a RW as well as a BRW, all time correlations are simply zero. 
%successive steps are uncorrelated.

 In a one dimensional walk, 
 two successive steps gives 
rise to four possible paths: 
 LR, LL, RL, RR. We investigate in detail the probabilities
of  these moves to gain further insight into the walks in the GLS as 
the correlations for successive time steps  are strongest. 
This correlation, $C(1)$,  is related to the probabilities $W$ of these moves; precisely,
$C(1) = W(RR) + W(LL) - W(LR) - W(RL) - s_{0}^2$. 

%Along with this, we also calculate 
%the correlations $P(1,1)$ and $P(-1,-1)$.  

\begin{figure}%[t]
\includegraphics[width=8.5cm,angle=0]{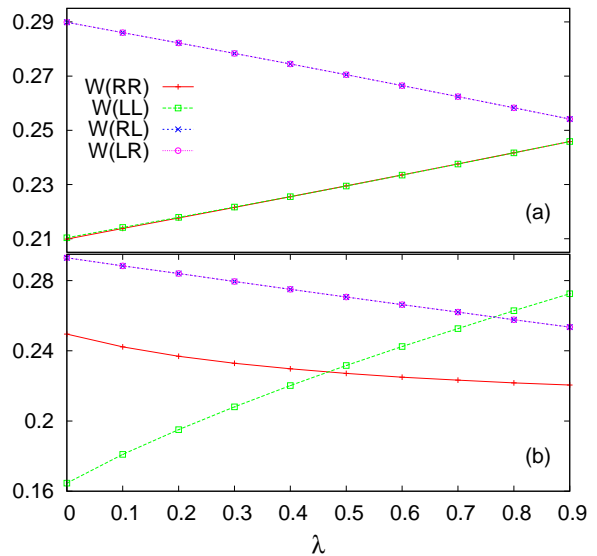}
%\includegraphics[width=6cm,angle=270]{prob_correlation_CCM.eps}
%\centering \includegraphics[width=6cm,angle=270]{correlation_time_mod.eps}
\caption{(Color online) The probabilities W(RR), W(LL), W(RL) and W(LR) are 
shown for (a) the CC and (b) the CCM models.
}
\label{corr-lambda}
\end{figure}

The results for both CC and CCM models are shown in Fig. \ref{corr-lambda}.
We  notice that irrespective of the value of $\lambda$, $W(RL) = W(LR)$, i.e., the tendency to change direction 
does not depend on the sequence of the steps taken.  
At the same time, we note that while for the CC walkers, there is 
also a symmetry $W(RR) = W(LL)$, for the CCM walkers, which have 
a bias, these two measures are unequal in general and become equal only at 
the ``bias-less'' point $\lambda^*$.
 
\begin{figure}%[t]
\includegraphics[width=8.5cm]{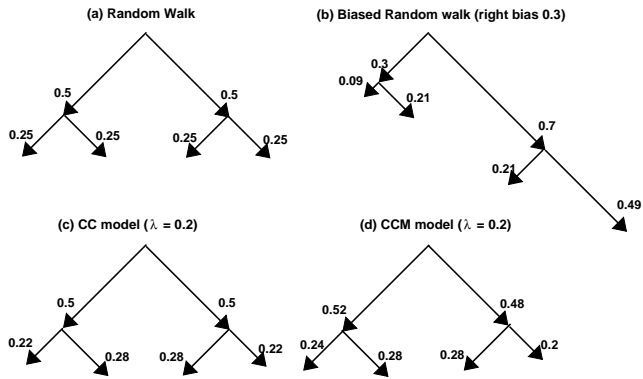}
\caption{
Typical movements of a  walker in two successive time steps:
 The four possible moves: LL, LR, RL, RR  are shown
where the step lengths  proportional
to the corresponding  probabilities (also shown). In all cases, probability of a LR move is equal to an RL move (symmetric).
In (a), the movement of a  random walker (RW) shows that it  moves with full symmetry.
(b) A biased random walker has equal probabilities for LR and RL but
unequal probabilities for the other two moves.
 (c) CC model:   Walkers have equal probabilities to move L and R in step 1 like RW,
but different probabilities for LR and LL (or RR)  in the next step.
However, there is symmetry for the RR and LL moves here.
(d) For the CCM model, the symmtery properties are similar to the  BRW  
 but to be noted is the fact that LR and RL moves occur with  probability $> 0.25$ (also true for CC walks).
Results are shown for $\lambda = 0.2$ for CC and CCM walks; qualitative feature is independent
of $\lambda$.}
\label{corr-schematic}
\end{figure}

From these detailed measures, it is now entirely clear how the CC walk differs 
from the RW and CCM from the BRW as illustrated  in Fig.~\ref{corr-schematic}.

\subsection{Understanding why direction change is preferred}
At this point it is apparent that in general in these kinetic 
exchange models, the tendency to make a gain and a loss 
in successive steps (in either order) is  independent of the saving 
feature of the CC and CCM models. In fact,   it is present with maximum 
probability in the CC model with $\lambda=0$ (i.e., the DY 
model) when agents do not save at all.

We therefore try to understand this feature from the point of view of the DY model
which has a simple, exactly known form  for the money distribution
by considering the transactions made in two successive steps.
We show below that indeed, for the DY case, it can be proved that 
the probability of direction changes is greater than $\frac{1}{2}$.

 In the DY model (and in fact in the CC model for any $\lambda$), in general, an agent gains/loses 
while interacting with a richer/poorer agent. This is because, if agent 1 with money $m_1$ interacts with agent 2 with 
money $m_2$,  after interaction,  agent 1 will have   money  $m^{\prime} = \epsilon(m_1+m_2)$.
On an average, if agent 1 gains, $(m_1+m_2)/2 > m_1$, or $m_2 > m_1$. 

To prove that the probability of direction changes is greater than $\frac{1}{2}$, we show that individually
$W(RL)$ and $W(LR)$ are greater than 1/4. Suppose an 
agent  had a gain in the first  step and ended up with money $m_g$.
Let  $W^\prime(LR)$ be the  conditional probability that the agent loses in the next step while interacting with 
another agent with money $m$, given that she/he  gained in the first step.
 This probability has to take care of two   factors:\\
(i) condition that  $m \leq m_g$,\\
 (ii) averaging over all possible $m_g$.

Using the money distribution function $P(m)  = \exp(-m)$ for the DY model (taking $T=M/N =1$), one gets
$$ W^\prime(LR) = \frac{\int_{m_{g}^0}^\infty P(m_g) dm_g  \int_0^{m_g} P(m) dm } {\int_{m_{g}^0}^\infty P(m_g) dm_g}$$
\begin{equation}
= 1-\frac{1}{2} \exp(-m_{g}^0). 
\end{equation}

The lower limit of the integral over $m_g$ is taken as $m_{g}^0$ and not zero since after a gain in the 
first step, the agent must have money greater than zero. $m_{g}^0$ may be considered to be  an arbitrary  lower bound.

Now  $W(LR) = W^\prime(LR)/2$ simply as we know that probability of R or L at the first step is just 1/2 \cite{Chatt-sen:2010}.
Therefore we find that  $W(LR) = [1-\frac{1}{2} \exp(-m_{g}^0)]/2 \geq 1/4$ independent of the value of 
$m_{g}^0$. 

In a similar manner, for the move RL, we  have an arbitrary upper bound $m_{l}^0$ for the money $m_l$ the agent ends up with 
after a loss in the first step such that 
$$W(RL) = \frac{1}{2}   \frac{\int_{0}^{m_{l}^0} P(m_l) dm_l  \int_{m_l}^\infty  P(m) dm } {\int_0^{m_{l}^0} P(m_l) dm_l}$$
\begin{equation} 
= \frac{1}{4} [1+\exp(-m_{l}^0)]. 
\end{equation}

 Obviously, $W(RL)$ is also greater than or equal to 1/4 
for all values of $m_{l}^0$ and therefore the sum  $W(LR) + W(RL) \geq \frac{1}{2}$.
Since $W(RL)$ and $W(LR)$ equal 1/4 for extremely improbable cases, one can conclude that  $W(LR) + W(RL) > \frac{1}{2}$ in general.

%********************************************************
%%%%%%%%%%%%%%%%%%%%%%%%%%%%%%%%%%%%%%%%%%%%%%%%%%%%%%%
What happens for CC and CCM models?
In the CC model, the conditional probability that an agent  loses after gaining   
depends on $\lambda$  through the money distribution function. Since its 
form is not exactly known, it is not possible to get exact results. 
 However,
for the CC model, there is a growing region for the money distribution
curve for small $m$ values and therefore the probability that agent 1 meets a poorer agent in the next step
is less probable  compared to the DY model and hence qualitatively it 
is understandable that $W(RL)$ or ($W(LR)$) will decrease with $\lambda$. 
At the same time it is true in the CC model also that the conditional probability $W^\prime$ is  twice of the probability 
$W$
of a LR  or a RL move as in the DY model, independent of $\lambda$ \cite{Chatt-sen:2010}.

In the CCM model, matters become  more complicated as 
the condition for gain/loss depends on the interacting agents' 
saving propensities. It was found in \cite{Chatt-sen:2010} that the probability of a gain is higher when one
interacts with an agent with larger $\lambda$. Since the average money of an agent increases with   $\lambda$
 \cite{Mohanty:2006} (in a nonlinear manner), this condition implies, once again, that a gain is more likely while 
interacting with a richer agent. Consequently, the same kind of logic 
holds good here: for a direction change to occur, an agent who got richer (poorer) in one step should  interact
with a poorer (richer) agent in the next. However, like the CC model,  the exact form of the money distribution is 
not known here. Moreover, 
 the  probabilities $W^\prime$ and $W$ for the LR and RL moves are not simply 
related in the CCM model.

%\subsection {Hurst exponent}
%
%So far we have only qualitatively found that an antipersistence effect exists for the 
%CC and CCM models. To get a quantitative measure of this, one can evaluate the Hurst exponent \cite{feder}
%which indicates whether a time series is correlated and in which way. For a completely random series, the
%Hurst exponent is expected to be equal to 1/2. A lesser value would imply antipersistence while 
%higher values indicate positive correlation, as in  biased random walk or persistent walks.
%Here we find that in general the hurst exponent is less than 0.5

\section{Comparisons with a simulated  walk (SW) of a single agent}
\label{sec:4}
%********************************************************
%%%%%%%%%%%%%%%%%%%%%%%%%%%%%%%%%%%%%%%%%%%%%%%%%%%%%%%%%%%%
In the previous section we found that the form of the money distribution
is responsible for the preference  of direction change in the 
gain loss space, leading to the result $f > 1/2$, although the actual amount of money lost/gained 
is ignored in the walk picture. This is true for all the
 kinetic exchange models considered,  whether
there is saving or not.
For CC, we had seen earlier that an agent gains when interacting with a richer agent, in case  of CCM the 
condition that an agent gains is less simple, involving the 
instantaneous money possessed by the two agents and their saving propensities 
 \cite{Chatt-sen:2010}. 
%, the choice of the second agent
%becomes an important factor giving rise to $f> 1/2$ in all cases.  
Hence we are led to believe that if a 
CCM/CC kind of walk is generated which  incorporates 
a probability of going to right or left according to 
the results obtained on an average but ignores the actual exchange of money 
taking place at every instant, 
 the result $f> 1/2$ will not be observed.
Such a walk for the CC is trivial, here all agents are 
identical and one only has to generate a walk which
has probability 1/2 of going either way making it completely identical 
to a RW. 

For the CCM, however, it is possible to generate a nontrivial single agent 
walk 
determined from the existing results. 
It was found in \cite{Chatt-sen:2010} that the probability of gain over loss
on an average for an agent with given saving propensity $\lambda_1$ while interacting with another agent whose saving propensity is  $\lambda_2$, has  the following  form:
\begin{equation}
\label{eq:pgpl}
\mathcal{P}_g - \mathcal{P}_l = {\rm const.} \frac{\lambda_2 -\lambda_1}{1.5+\lambda_1 + \lambda_2}.
\end{equation}
The constant  turns out to be very close to  0.345.
Eq \ref{eq:pgpl} suggests that at each step, a tagged walker with saving $\lambda_1$ will  to move left/right with a probability which depends on $\lambda_2$ as well. This probability at each step  can be
calculated easily from the above equation once $\lambda_1$ and $\lambda_2$ values are known 
and using the fact that $\mathcal{P}_g + \mathcal{P}_l =1$.  
It is therefore possible to generate a walk
for a single  agent with given $\lambda_1$, assuming
that  at each step it interacts with another agent of  randomly chosen $\lambda_2$ to give the probability of
movement to right/left at that instant.
Thus in this walk, 
    the money distribution
function does not enter the picture at all and at the same time 
the probability of a move towards any direction is not fixed.

It is interesting to compare the results of this simulated walk with the original multiagent CCM walk.
We find that
in fact the effective $p$ values are almost identical. 
For the SW, we can extract an effective $p$ in two ways - first is as usual by calculating the slope (section IIA), and 
secondly by taking the average value of the probability $\mathcal{P}_g$ (to move right)
generated  for all times - these two values are very close. Only 
the value obtained using the slopes have been  shown in Fig  \ref{p-slope} along with the results for the CCM walk. 

\begin{figure}%[t]
\includegraphics[width=6cm,angle=270]{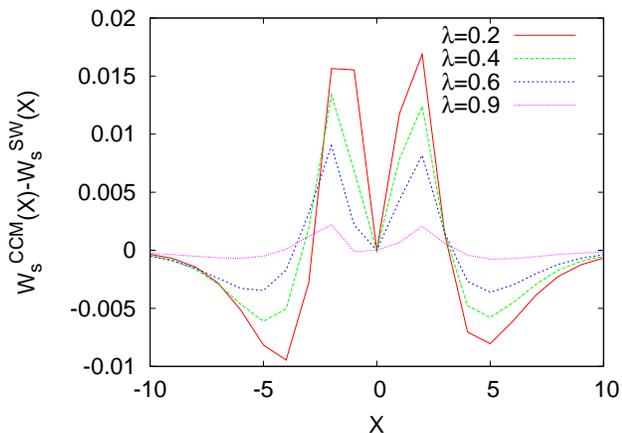}
\caption{
(Color online) Difference between the probabilities $W_{s}(X)$ calculated from the 
CCM model and the SW are plotted against $\lambda$}
\label{compare-PX}
\end{figure}

\begin{figure}%[t]
\centering \includegraphics[width=6cm,angle=270]{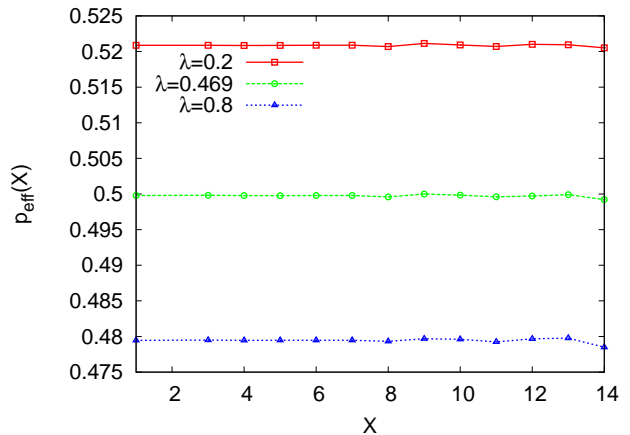}
\caption{(Color online) Variation of $p_{eff}(X)$ against $X$ for different $\lambda =  0.2, 0.469, 0.8$ for the 
simulated walk  (SW).  
}
\label{p-vs-X-SW}
\end{figure}

However, when we calculate $W_{s}(X)$ for the simulated walks, it turns out 
that these are not at all comparable to the CCM (Fig \ref{compare-PX}). 
There are two interesting features to be noted here: the probabilities for small $X$ is larger for the CCM
model and the magnitude of differences decrease with $\lambda$. Both these results can be explained from the 
antipersistence effect present in  the CCM model. Here the increased number of direction 
changes results in a larger value of  $W_s(X) $  for small $X$ and the fact that antipersistence 
effect decreases with $\lambda$ makes the CCM and SW models more similar as $\lambda$ increases.
We also note that the $p$ values extracted from the ratio $W_{s}(X)/W_{s}(-X)$ are indeed independent of $X$ 
(Fig.~\ref{p-vs-X-SW})
which is expected for a BRW. So the simulated single agent walk is like a conventional BRW,
compared to the CCM where the $p$ values have a dependence on $\lambda$ as well as the number of steps $X$
(Fig.~\ref{p-vs-X-CCM}).

When $f$, the fraction of direction changes is calculated for the simulated walk, we find that it is  less than  
$\frac{1}{2}$ for
all values of $\lambda$, and very close to $\frac{1}{2}$ at $\lambda^* \simeq 0.469$ 
(Fig.~\ref{direc-change}). 
%All the results obtained for the SW are thus   compatible with the fact that 
% the walker performs a biased random walk  unlike  the agents in  the original multiagent CCM model. 

The simulated walk of a single agent  once again 
shows the presence of a $\lambda^* \approx 0.47$ where the walk becomes 
bias-less,  but otherwise shows features which are identical to those of  BRW. 
This is consistent with the conjecture that the choice of the second agent 
is crucial where the money distribution form plays a significant role.

%********************************************************

%%%%%%%%%%%%%%%%%%%%%%%%%%%%%%%%%%%%%%%%%%%%%%%%%%%%%%%%%%%%%%%%5

\section{Summary and Concluding remarks}

In this paper, we explored the nature of transactions made in some kinetic exchange models of wealth 
distribution in  depth. 
Using an equivalent picture of a one dimensional walk in an abstract space for gains
and losses, it is found that 
there is a tendency of individuals to make a gain immediately after a loss
and vice versa. This so called antipersistence effect is in fact 
compatible with 
human psychology; one can afford to incur a loss after a gain and will try to
gain after suffering a loss.

Moreover we find that if there is no saving factor, this 
effect is maximum and decreases with saving.  This is perhaps  in tune  with 
the  human feeling of security   associated with the saving factor.
In the CCM model, where the saving propensity is randomly distributed, the antipersistence effect occurs with a simultaneous 
bias which too depends on $\lambda$. Whether the two features are 
correlated is a matter of future work.

 The antipersistence effect makes the CCM walkers different from biased random walkers 
such that the $\langle X \rangle$
values (Eq. \ref{avX+}) become drastically different as shown in Section II. However, the quantity $\langle X \rangle_{-}$  (Eq. \ref{avX-})   
is apparently not affected. This is 
because the antipersistence effect is symmetric 
(as $W(RL) = W(LR)$)  
 and  it cancels out in  $\langle X \rangle_{-}$,
    while the asymmetry due to the bias remains and one gets good agreement with BRW results using this measure.

%In reality, the CCM walkers are
% not ordinary biased random walkers.
%The CC walkers,  on the other hand, have no bias, but once again
%they are not simple random walkers because of the antipersistence effect. 

There is also no antipersistence effect in the simulated walk of a 
single agent of CCM type, as is evident from the comparative behaviour of 
the probability of direction reversal of the two models (Fig \ref{direc-change}) 
when the actual money exchange process is not taken under consideration.
This shows that  although we consider only gains and losses for generating the walk in 
a GLS, the form of the money distribution function is   
crucial. 

Just as in the CCM model, where the antipersistence effect makes it distinct from an ordinary biased random walk,
in the CC model, the antipersistence effect makes it different from an ordinary random walk.
The antipersistence effect for the CC and CCM models manifests itself through 
the quantities $W(LR)$ and $W(RL)$, both of which are greater than $\frac{1}{4}$. In fact, 
as is evident from Fig \ref{corr-lambda}, $W(LR)$ (or $W(RL)$) values are numerically very close to each other for CC and CCM 
models as a function of $\lambda$. The difference in CC and CCM walks thus turns out to be 
simply the  
presence of a bias in 
the latter. We believe that this bias appears as a result of the   small positive correlations remaining 
at large times in the CCM model. 

The different calculations made in this work shows once again 
that an agent with $\lambda = \lambda^* \simeq 0.469$ 
in the CCM model is identical to a CC walker.
An analytical value of $\lambda^*$ may be obtained if the exact 
form of the money distribution curve is known using the fact that 
the overall gains and losses made by the agent with  $\lambda^*$ 
are exactly equal.

\begin{acknowledgements}
The authors thank Deepak Dhar and S. S. Manna
for some useful comments and discussions, and
Soumyajyoti Biswas for critical reading of the manuscript.
SG acknowledges financial support from CSIR (Grant no. 09/028(0762)/2010-EMR-I). 
PS acknowledges financial support from DST grant and partial computational support from UPE project.
\end{acknowledgements}

%Points to be noted here
%
%1. CCM -Not a conventional BRW; \\
%2. Even at $\lambda^*$, no RW behaviour.\\
%3. CC walk - not a RW 
%4. The question of variance  scaling as $t^2$ needs to be addressed here perhaps. Cannot be gaussian
%5. perhaps the issue of diverging time scale also mentioned here briefly rather than having a section on it
%
%
%%%%%%%%%%%%%%%%%%%%%%%%%%%%%%%%%%%%%%%%%%%%%%%%%%%%%%%%%%%%%%%%5

%%%%%%%%%%%%%%%%%%%%%%%%%%%%%%%%%%%%%%%%%%%%%%%%%%%%%%%%%%%%%%%%%%%%

%%%%%%%%%%%%%%%%%%%%%%%%%%%%%%%%%%%%%%%%%%%%%%%%%%%%%%%%%%%%%%%%%%%
\end{document}